\documentclass[aps,floatfix,footinbib, twocolumn, superscriptaddress,amsmath,amssymb,amsfonts,notitlepage]{revtex4-1} 

\usepackage[T1]{fontenc}
\usepackage{lmodern} 
\usepackage{color}
\usepackage{bbold}
\usepackage{dsfont}
\usepackage{graphicx}
\usepackage[caption=false]{subfig}
\usepackage[colorlinks]{hyperref}
\usepackage[bottom]{footmisc}
\usepackage{enumerate}
\usepackage{float}
\usepackage{mathtools}

\usepackage{empheq}
\usepackage{tikz}
\usepackage{fancyhdr}
\usepackage[normalem]{ulem}
\DeclareMathAlphabet\mbc{OMS}{cmsy}{b}{n}
\usepackage{balance}

\begin{document}

\global\long\def\eqn#1{\begin{align}#1\end{align}}
\global\long\def\vec#1{\overrightarrow{#1}}
\global\long\def\ket#1{\left|#1\right\rangle }
\global\long\def\bra#1{\left\langle #1\right|}
\global\long\def\bkt#1{\left(#1\right)}
\global\long\def\sbkt#1{\left[#1\right]}
\global\long\def\cbkt#1{\left\{#1\right\}}
\global\long\def\abs#1{\left\vert#1\right\vert}
\global\long\def\cev#1{\overleftarrow{#1}}
\global\long\def\der#1#2{\frac{{d}#1}{{d}#2}}
\global\long\def\pard#1#2{\frac{{\partial}#1}{{\partial}#2}}
\global\long\def\re{\mathrm{Re}}
\global\long\def\im{\mathrm{Im}}
\global\long\def\dd{\mathrm{d}}
\global\long\def\ddd{\mathcal{D}}

\global\long\def\avg#1{\left\langle #1 \right\rangle}
\global\long\def\mr#1{\mathrm{#1}}
\global\long\def\mb#1{{\mathbf #1}}
\global\long\def\mc#1{\mathcal{#1}}
\global\long\def\tr{\mathrm{Tr}}
\global\long\def\dbar#1{\Bar{\Bar{#1}}}

\global\long\def\nth{$n^{\mathrm{th}}$\,}
\global\long\def\mth{$m^{\mathrm{th}}$\,}
\global\long\def\non{\nonumber}

\newcommand{\orange}[1]{{\color{orange} {#1}}}
\newcommand{\teal}[1]{{\color{teal} {#1}}}
\newcommand{\cyan}[1]{{\color{cyan} {#1}}}
\newcommand{\blue}[1]{{\color{blue} {#1}}}
\newcommand{\yellow}[1]{{\color{yellow} {#1}}}
\newcommand{\green}[1]{{\color{green} {#1}}}
\newcommand{\red}[1]{{\color{red} {#1}}}
\newcommand{\pbb}[1]{{\textcolor{teal}{[PBB: #1]}}}

\global\long\def\todo#1{\orange{{$\bigstar$ \cyan{\bf\sc #1}}$\bigstar$} }
\newcommand{\ks}[1]{{\textcolor{teal}{[KS: #1]}}}

\newcommand{\bd}{{\bf d}}      
\newcommand{\bv}{{\bf v}}
\newcommand{\hbp}{\hat{\bp}}
\newcommand{\hbx}{\hat{\bx}}
\newcommand{\hq}{\hat{q}}
\newcommand{\hp}{\hat{p}}
\newcommand{\ha}{\hat{a}}
\newcommand{\had}{{a}^{\dag}}
\newcommand{\ad}{a^{\dag}}
\newcommand{\hsig}{{\hat{\sigma}}}
\newcommand{\nt}{\tilde{n}}
\newcommand{\itf}{\sl}
\newcommand{\eps}{\epsilon}
\newcommand{\bsig}{\pmb{$\sigma$}}
\newcommand{\beps}{\pmb{$\eps$}}
\newcommand{\bmu}{\pmb{$ u$}}
\newcommand{\balpha}{\pmb{$\alpha$}}
\newcommand{\bbeta}{\pmb{$\beta$}}
\newcommand{\bgamma}{\pmb{$\gamma$}}
\newcommand{\bu}{{\bf u}}
\newcommand{\bpi}{\pmb{$\pi$}}
\newcommand{\bSig}{\pmb{$\Sigma$}}
\newcommand{\be}{\begin{equation}}
\newcommand{\ee}{\end{equation}}
\newcommand{\bea}{\begin{eqnarray}}
\newcommand{\eea}{\end{eqnarray}}
\newcommand{\sss}{_{{\bf k}\lambda}}
\newcommand{\ssss}{_{{\bf k}\lambda,s}}
\newcommand{\dip}{\langle\sigma(t)\rangle}
\newcommand{\dipp}{\langle\sigma^{\dag}(t)\rangle}
\newcommand{\sig}{{{\sigma}}}
\newcommand{\sigd}{{\sigma}^{\dag}}
\newcommand{\sigz}{{\sigma_z}}
\newcommand{\ra}{\rangle}
\newcommand{\la}{\langle}
\newcommand{\om}{\omega}
\newcommand{\Om}{\Omega}
\newcommand{\pa}{\partial}
\newcommand{\bR}{{\bf R}}
\newcommand{\bx}{{\bf x}}
\newcommand{\br}{{\bf r}}
\newcommand{\bE}{{\bf E}}
\newcommand{\bH}{{\bf H}}
\newcommand{\bB}{{\bf B}}
\newcommand{\bP}{{\bf P}}
\newcommand{\bD}{{\bf D}}
\newcommand{\bA}{{\bf A}}
\newcommand{\bek}{{\bf e}\rmk}
\newcommand{\rmk}{_{{\bf k}\lambda}}
\newcommand{\rk}{_{{\bf k}_1{\lambda_1}}}
\newcommand{\rkk}{_{{\bf k}_2{\lambda_2}}}
\newcommand{\rkz}{_{{\bf k}_1{\lambda_1}z}}
\newcommand{\rkkz}{_{{\bf k}_2{\lambda_2}z}}
\newcommand{\bsij}{{\bf s}_{ij}}
\newcommand{\bk}{{\bf k}}
\newcommand{\bp}{{\bf p}}
\newcommand{\epso}{{1\over 4\pi\eps_0}}
\newcommand{\BB}{{\mathcal B}}
\newcommand{\AAA}{{\mathcal A}}
\newcommand{\NN}{{\mathcal N}}
\newcommand{\mm}{{\mathcal M}}
\newcommand{\RR}{{\mathcal R}}
\newcommand{\bS}{{\bf S}}
\newcommand{\bL}{{\bf L}}
\newcommand{\bJ}{{\bf J}}
\newcommand{\bI}{{\bf I}}
\newcommand{\bF}{{\bf F}}
\newcommand{\bsub}{\begin{subequations}}
\newcommand{\esub}{\end{subequations}}
\newcommand{\baline}{\begin{eqalignno}}
\newcommand{\ealine}{\end{eqalignno}}
\newcommand{\isat}{{I_{\rm sat}}}
\newcommand{\Is}{I^{\rm sat}}
\newcommand{\Ip}{I^{(+)}}
\newcommand{\Imm}{I^{(-)}}
\newcommand{\Inu}{I_{\nu}}
\newcommand{\bInu}{\overline{I}_{\nu}}
\newcommand{\bN}{\overline{N}}
\newcommand{\qnu}{q_{\nu}}
\newcommand{\oqn}{\overline{q}_{\nu}}
\newcommand{\qsat}{q^{\rm sat}}
\newcommand{\Iout}{I_{\nu}^{\rm out}}
\newcommand{\topt}{t_{\rm opt}}
\newcommand{\crr}{{\mathcal{R}}}
\newcommand{\cE}{{\mathcal{E}}}
\newcommand{\cH}{{\mathcal{H}}}
\newcommand{\epsoo}{\epsilon_0}
\newcommand{\ombar}{\overline{\om}}
\newcommand{\cEp}{{\mathcal{E}}^{(+)}}
\newcommand{\cEm}{{\mathcal{E}}^{(-)}}
\newcommand{\bvv}{\tilde{\bv}}
\newcommand{\pr}{^{\prime}}
\newcommand{\dpr}{^{\prime\prime}}
\newcommand{\hk}{\hat{\bk}}
\newcommand{\hn}{\hat{\bf n}}
\newcommand{\Ep}{{\bf E}^{(+)}}
\newcommand{\Em}{{\bf E}^{(-)}}
\newcommand{\gd}{g^{\dag}}

\title{Scalar QED Model for Polarizable Particles in Thermal Equilibrium or in Hyperbolic Motion in Vacuum}
\author{Kanu Sinha}
\affiliation{College of Optical Sciences and Department of Physics, University of Arizona, Tucson, Arizona 85721 USA}
\email{kanu@arizona.edu}
\author{Peter W. Milonni}
\affiliation{Department of Physics and Astronomy, University of Rochester, Rochester, NY 14627 USA}
\email{peter\_milonni@comcast.net}

\begin{abstract}
    We consider a scalar QED model for the frictional force and the momentum fluctuations of a polarizable particle in thermal equilibrium with radiation or in hyperbolic motion in a vacuum. In the former case the loss of particle kinetic energy due to the frictional force is compensated by the increase in kinetic energy associated with the momentum diffusion, resulting in the Planck distribution when it is assumed that the average kinetic energy satisfies the equipartition theorem. For hyperbolic motion in vacuum the frictional force and the momentum diffusion are similarly consistent with a thermal equilibrium at the Davies-Unruh temperature. The quantum fluctuations of the momentum imply that it is only the {\it average} acceleration that is constant when the particle is subject to a constant applied force. 
\end{abstract}
\maketitle

\section{Introduction}
The question of whether there is radiation from a uniformly accelerated oscillator in vacuum continues to be of interest \cite{rfo}. In this paper we revisit the question within the context of a 2D scalar QED model \cite{raine, rfo1}. Our approach differs from previous work with this model in that we focus on the force on the particle and its momentum fluctuations as it exchanges energy with the field. Since it is well established that the particle will perceive itself to be in a field with a Planck distribution at a temperature $T=\hbar a/2\pi k_Bc$, where $a$ is the acceleration and $k_B$ is Boltzmann's constant \cite{davies, unruh}, we first consider, within the 2D scalar QED model, a polarizable particle in equilibrium with thermal radiation in the absence of any external force that would accelerate the particle. We follow the original approach of Einstein and Hopf \cite{einhopf} and later Einstein \cite{ein} in which it is shown that the frictional force and the momentum diffusion experienced by the particle are consistent with the Planck distribution for the field. The 2D scalar QED model exhibits the basic physics of the problem while permitting some simplification compared to the complete QED theory \cite{milton,sinha}. 

Essentially the same approach in the case of a uniformly accelerated particle in vacuum yields the expected result: the particle undergoes a frictional force and momentum fluctuations as if it were in thermal equilibrium at the temperature $T=\hbar a/2\pi k_Bc$. The result for the momentum fluctuations seems particularly interesting in the way that the Bose--Einstein form of the fluctuations emerges. The coupling of the particle to the quantum field results in quantum fluctuations about the its classical hyperbolic trajectory. These fluctuations are superposed on an otherwise uniformly accelerated motion.

In the following section we present the basic equations of motion needed for our analysis. In Section \ref{sec:friction} we derive, based on the Lorentz transformation of the spectral energy density of the field, the frictional force on a polarizable particle in a field of blackbody radiation, and in Section \ref{momentum} we derive the momentum fluctuations experienced by such a particle. The expressions obtained for the frictional force and the momentum fluctuations are shown to be consistent with a Planck distribution for the spectral energy density of the field. In Section \ref{hyperbolic} we extend these considerations to the case of hyperbolic motion in a vacuum. Our results are summarized in Section \ref{summary}.

\section{Hamiltonian and Equations of Motion}

\begin{figure}[t]
    \centering
    \includegraphics[width = 0.35\textwidth]{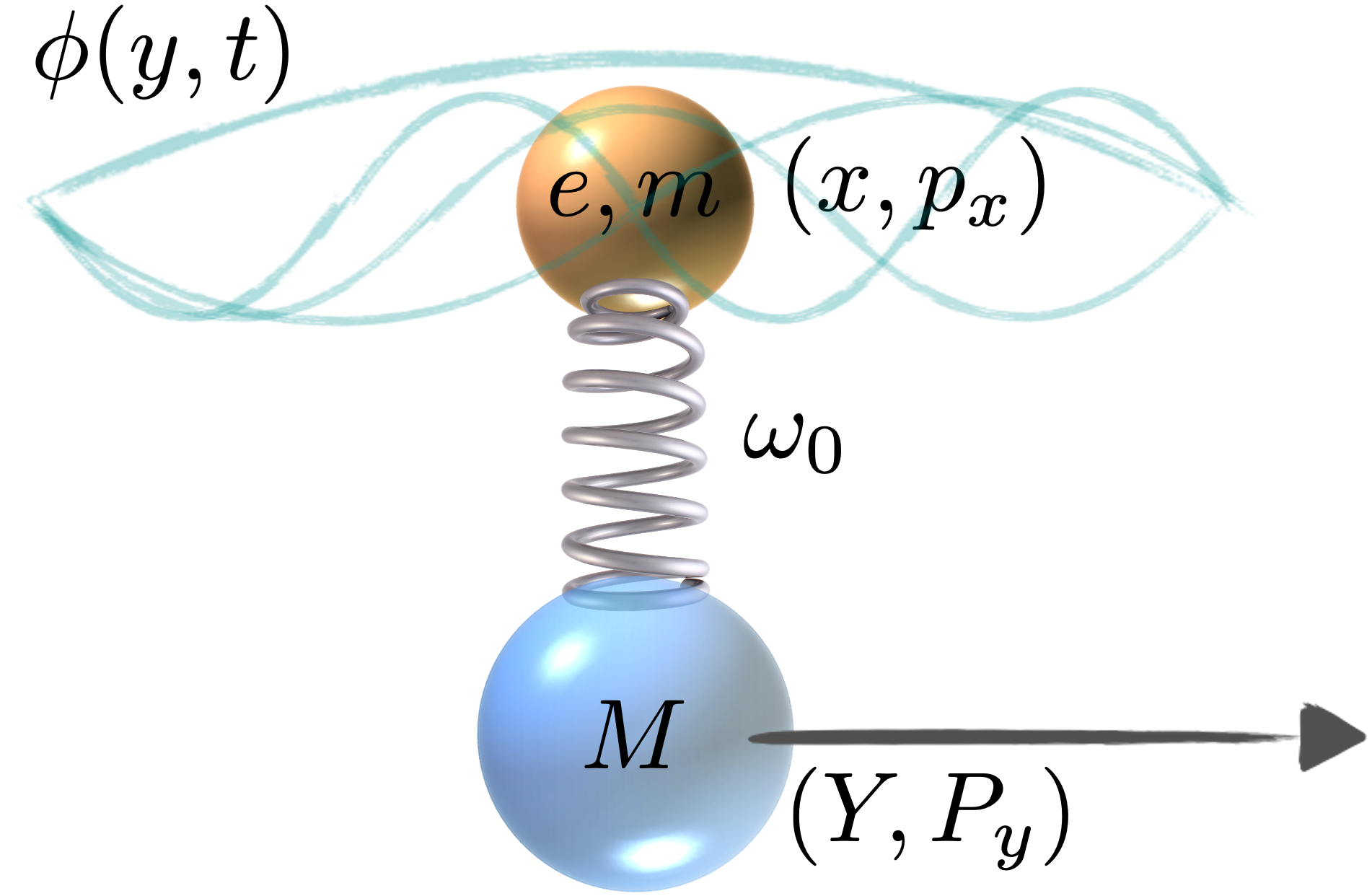}
    \caption{Schematic representation of a point particle of mass $ M$ and harmonically bound oscillator with a charge $ e$ and mass $ m$. The charged oscillator moves along the $ x $-axis with a frequency $ \omega_0 $ and interacts with a scalar field $\phi(y, t)$, where $ Y$ is the center-of-mass position of the point particle.}
    \label{Fig:Sch}
\end{figure}

We consider a point particle  interacting with a 1 + 1D scalar field $\phi(y,t)$ via a harmonically bound charged oscillator, as shown in Fig.~\ref{Fig:Sch}. The Lagrangian of the system is
\eqn{
\mc{L} = \frac{1}{2}M \dot{Y}^2 + \frac{1}{2}m \dot{x}^2 - \frac{1}{2}m \omega_0^2 x^2+ \int \dd y \sbkt{\bkt{\partial_t \phi(y,t)}^2 \right.&\non\\
 \left.- \bkt{\partial_y \phi(y,t)}^2 + e \phi(y,t) \dot x \delta (y - Y)}.&
}
We denote the center-of-mass position of the particle by $ Y(t)$, the amplitude of the charged oscillator degree of freedom by $ x(t) $. The charged oscillator interacts with the field, and is constrained to be located at the center-of-mass position $Y(t)$ of the particle~\cite{microOM}.   Defining the canonical conjugate momenta associated with the various degrees of freedom as $P_y = M \dot{Y}$, 
$p_x = m\dot x + e\phi(Y,t)$ and 
$\Pi (y,t) =  \dot{\phi} (y,t)$, we obtain the corresponding  Hamiltonian in the minimal coupling form as
\eqn{
H=P_y^2/2M+\frac{1}{2}m\om_0^2x^2+\frac{1}{2m}\sbkt{p_x-e\phi (Y, t)}^2&\non\\
+\sum_k\hbar\om_k\ad_ka_k,&}
with
\be\label{eq:phiyt}
\phi(y,t)=\sum_kC_k\big[a_k(t)e^{iky}+\ad_k(t)e^{-iky}\big];\\
C_k=\bkt{\frac{2\pi\hbar}{\om_kL}}^{1/2}
\ee
  $a_k$ and $\ad_k$ are the bosonic annihilation and creation operators following the canonical commutation relations $[a_k,\ad_{k\pr}]=\delta_{kk\pr}$  and the quantization volume is of length $L$. (We omit the speed of light $c$ in our equations.) From the Heisenberg equations of motion that follow from this Hamiltonian we obtain the following equations for the operators $x,\phi_0$, and (see Appendix A)
\be\ddot{x}+2\beta\dot{x}+\om_0^2x=-\frac{e}{m}\frac{\pa\phi_0(Y,t)}{\pa t},\label{eq1}\ee
\be\phi_0(y,t)=\sum_kC_k\big[a_{k}e^{-i(\om_kt-ky)}+\ad_{k}e^{i(\om_kt-ky)}\big],\label{eq2}\ee
\be\dot{P}_y=\frac{e}{2}\bkt{\dot{x}\frac{\pa\phi_0(Y, t)}{\pa y}+\frac{\pa\phi_0(Y, t)}{\pa y}\dot{x}}
\label{eq3}.\ee
$\phi_0(y,t)$ is the source-free part of $\phi(y,t)$; the source (``radiation reaction") part of $\phi(y,t)$ results in the damping force $-2\beta m\dot{x}$ in equation (\ref{eq1}). $\beta$ is found without approximation to be $\pi e^2/m$, consistent with the commutation relation $[x(t),m\dot{x}(t)]=i\hbar$ maintained by the formal solution of equation (\ref{eq1}) \cite{pwm1}.

Equations (\ref{eq1}) and (\ref{eq2}), together with the equation for the radiation field $\phi(y,t)$, are the basis for the analyses by Raine et al. \cite{raine} and Ford and O'Connell \cite{rfo1} and their conclusion that an oscillator in hyperbolic motion does not radiate. We focus instead on equation (\ref{eq3}). In other words, we focus not on the radiation field but on the force acting on the particle. In the case of hyperbolic motion, this force results in fluctuations superposed on the otherwise uniform acceleration.

For the linear response of $x(t)$ to $\pa\phi_0(Y,t)/\pa t$ we define the
polarizability
\be
\alpha(\om)=\frac{e^2/m}{\om_0^2-\om^2-2i\beta\om}
\label{pol}
\ee
such that the steady-state expression for $ex(t)$ for the particle at rest is
\be
ex(t)=i\sum_kC_k\om_k[\alpha(\om_k)a_{k}e^{-i\om_kt}-\alpha^*(\om_k)\ad_{k}e^{i\om_kt}],
\label{momm7}
\ee
where we have assumed without loss of generality that $ Y = 0$ specifies the particle coordinate.
\section{Frictional Force}\label{sec:friction}
A polarizable particle moving in a blackbody field experiences a drag force, as shown nonrelativistically by Einstein and Hopf \cite{einhopf} and Einstein \cite{ein}. To obtain a relativistic expression for the drag force we can proceed as in Reference \cite{milton} or Reference \cite{sinha}, for instance. Here we take a different, more heuristic approach based on the Lorentz transformation of the Planck spectrum in the 2D scalar QED model. Our starting point is the expression for the free-space energy density of the 2D massless scalar field:
\begin{align}
E=\frac{1}{2}\sbkt{\bkt{\frac{\pa\phi}{\pa t}}^2+\bkt{\frac{\pa\phi}{\pa y}}^2},    
\end{align}
which is proportional to $\om^2$ for a mode of frequency $\om$. In a frame S$\pr$ moving with velocity $v$ with respect to the frame in which the energy density is $E$, the energy density $E\pr\propto {\om\pr}^2$, with
\begin{eqnarray}
\om&=&\gamma{\om\pr}(1+v)  \nonumber\\
E&=&\gamma^2(1+v)^2E\pr, \ \ \ \gamma=(1-v^2)^{-1/2}.
\end{eqnarray}
We define the spectral energy density $\rho(\om)$ such that $\rho(\om)d\om$ is the energy per unit length (or ``volume") in the frequency interval $[\om,\om+d\om]$. In the moving frame $\rho(\om,\hat{\bk})\rightarrow 
\rho\pr(\om\pr,\hat{\bk})$, 
\begin{align}
\rho\pr(\om\pr,\hat{\bk})d\om\pr=\frac{E\pr}{E}\rho(\om,\hat{\bk})d\om=\frac{1}{\gamma^2(1+v)^2}\rho(\om,\hat{\bk})d\om 
\end{align}
for a plane wave of frequency $\om$ propagating in the direction of the unit vector $\hat{\bk}$, and therefore
\begin{align}
 \rho\pr(\om\pr,\hat{\bk})  =\frac{1}{\gamma^2(1+v)^2}\rho(\om,\hat{\bk})\frac{d\om}{d\om\pr}=\frac{1}{\gamma(1+v)}\rho(\om,\hat{\bk}).
\end{align}
For plane waves propagating in the direction parallel to the velocity ${\bf v}$, 
\begin{align}
\rho\pr(\om\pr)=\frac{\rho[\gamma\om\pr(1+v)]}{\gamma(1+v)}.
\label{wed191}
\end{align}

The energy eigenvalues for the free field are ($\om_k=|k|$)
\begin{align}
\sum_k\hbar\om_k\bkt{n(\om_k)+\frac{1}{2}}=&\frac{L}{2\pi}\int_{-\infty}^{\infty}dk\hbar\om\bkt{n(\om)+\frac{1}{2}}\\
 &\equiv L\int_0^{\infty}d\om\rho(\om),   
\end{align}
where the spectral energy density 
\be
\rho(\om)=\frac{\hbar\om}{\pi}\bkt{n(\om)+\frac{1}{2}}. 
\label{wed192}
\ee
From (\ref{wed191}) and (\ref{wed192}),
\be
\om\pr\bkt{n(\om)+\frac{1}{2}}=\om\pr\sbkt{n\bkt{\gamma\om\pr(1+v)}+\frac{1}{2}},
\ee
and so the Lorentz-transformed spectral energy density is
\be\rho\pr(\om\pr)=\frac{\hbar\om\pr}{\pi}\sbkt{n\bkt{\gamma\om\pr(1+v)}+\frac{1}{2}}.
\label{novv5}
\ee
In the case of thermal equilibrium at temperature $T$,
\begin{align}
n\bkt{\gamma\om\pr(1+v)}+\frac{1}{2}=\frac{1}{2}\coth\sbkt{\frac{\gamma\hbar\om\pr}{2k_BT}(1+v)}
\end{align}
and
\begin{align}
\rho\pr(\om\pr)=\frac{\hbar\om\pr}{2\pi}\coth\sbkt{\frac{\gamma\hbar\om\pr}{2k_BT}(1+v)}.   
\end{align}
This is the 2D version of the Lorentz-transformed Planck spectrum obtained in earlier work \cite{pauli,rfo2}.

The force on the particle may be expressed as
\eqn{
\mc{F}=& \frac{1}{2}\int_0^{\infty}d\om R(\om)\rho\pr(\om)\non\\
=& \frac{\hbar}{2\pi}\int_0^{\infty}d\om R(\om)\om\sbkt{n\bkt{\gamma\om(1+v)}+\frac{1}{2}},
\label{nov5}
}
where $R(\om)$ is the rate at which the field loses energy (and momentum $p=E$) in a given direction ($k>0$ or $k<0$), this momentum being taken up by the particle. Perhaps the simplest way to obtain an expression for $R(\om)$
is to start from the wave equation for $\phi$ \cite{microOM}:
\be
\frac{\pa^2\phi}{\pa y^2}-\frac{\pa^2\phi}{\pa t^2}=4\pi e\dot{x}\delta(y-Y),
\ee
which has the same form as the wave equation for an electric field propagating in one direction in a medium with number density $N \ (=\delta(y-Y)$ here) of particles with polarizability $\alpha(\om)$. In that case the energy dissipation rate $R(\om)=2k_I(\om)\cong 4\pi N\om\alpha_I(\om)$ in the dilute-medium approximation \cite{pwmjhe}.

In the present case we have one particle in our volume $L$, implying $R(\om)=4\pi\om\alpha_I(\om)$ and 
\be
\mc{F}=2\hbar\int_0^{\infty}d\om\om^2\alpha_I(\om)\sbkt{n\bkt{\gamma\om(1+v)}+\frac{1}{2}}.
\ee
This expresses the average force in terms of the linear response of the particle to a broadband field with Doppler-shifted frequencies $\om\sqrt{(1+v)/(1-v)}.$
Including field modes with $\bk$ either parallel or anti-parallel to the particle velocity ${\bf v}$, we obtain the net force
\eqn{
\mc{F}=&2\hbar\int_0^{\infty}d\om\om^2\alpha_I(\om)\sbkt{n\bkt{\gamma\om(1+v)}-n\bkt{\gamma\om(1-v)}}\non\\
\cong& 4\hbar v\int_0^{\infty}d\om\om^3\alpha_I(\om)\frac{\pa n}{\pa\om}\nonumber\\
=&-4\pi v\int_0^{\infty}d\om\om\alpha_I(\om)\sbkt{\rho-\om\frac{\pa\rho}{\pa\om}}.
\label{mom6}
}
In the full QED theory, in contrast, the nonrelativistic approximation to the friction force is \cite{remark1}
\be
\mc{F}=-4\pi v\int_0^{\infty}d\om\om\alpha_I(\om)\sbkt{\rho-\frac{\om}{3}\frac{\pa\rho}{\pa\om}},
\label{momm1}
\ee
with $\rho(\om)=\hbar\om^3n(\om)/\pi^2c^3$. It is noteworthy that Einstein and Hopf \cite{einhopf} and Einstein \cite{ein} obtain $\rho(\om)-\frac{\om}{3}\frac{\pa\rho}{\pa\om}$ in their formula for the friction force from essentially just kinematic effects involving Doppler shifts and aberration.

\section{Momentum Fluctuations and Thermal Equilibrium}\label{momentum}
From Eq. (\ref{eq3}),
\begin{widetext}
\eqn{
P_y(t)&=&i\sum_K\sum_kC_KC_k\om_K^2k\sbkt{\alpha(\om_K)a_Ka_kg_{Kk}(t)-\alpha(\om_K)a_K\ad_kf_{Kk}(t)+\alpha^*(\om_K)\ad_Ka_kf_{Kk}^*(t)-
\alpha^*(\om_K)\ad_K\ad_kg_{Kk}^*(t)},
\label{mom1}
}
\end{widetext}
where
\eqn{
f_{Kk}(t)=&\int_0^td\tau e^{-i(\om_K-\om_k)\tau} \ \ \ \ {\rm and}\non\\
g_{Kk}(t)=&\int_0^td\tau e^{-i(\om_K+\om_k)\tau}.
}
The calculation of momentum fluctuations $\avg{\Delta P_y^2(t)} = \la P_y^2(t)\ra$  is simplified if we assume that $\phi_0$ and its derivatives may be regarded as Gaussian random processes. This allows us to treat the $K$ and $k$ modes ($K\neq k$) in (\ref{mom1}) as uncorrelated \cite{einhopf2}. We further simplify by dropping terms involving $g_{Kk}(t)$, since their nonresonant character results in no contribution to $\la P_y^2(t)\ra$. Then
\begin{widetext}
    
\eqn{
&\la P_y^2(t)\ra=\sum_K\sum_kC_K^2C_k^2\om_K^4k^2|\alpha(\om_K)|^2\Big[\la a_K\ad_K\ra\la a_k\ad_k\ra+\la\ad_Ka_K\ra\la\ad_ka_k\ra\Big]|f_{Kk}(t)|^2\nonumber\\
\rightarrow& \Big(\frac{L}{2\pi}\Big)^2\int_{-\infty}^{\infty}dK\Big(\frac{2\pi\hbar}{\om_KL}\Big)\om_K^4|\alpha(\om_k)|^2\int_{-\infty}^{\infty}dk\Big(\frac{2\pi\hbar}{\om_kL}\Big)k^2\Big([n(\om_K)+1]n(\om_k)+n(\om_K)[n(\om_k)+1]\Big)
\frac{\sin^2\frac{1}{2}(\om_K-\om_k)t}{(\om_K-\om_k)^2/4}\nonumber\\
\cong&4\hbar^2\int_0^{\infty}d\om\om^4|\alpha(\om)|^2\int_0^{\infty}d\om\pr[2n(\om)n(\om\pr)+n(\om)+n(\om\pr)]2\pi t\delta(\om-\om\pr)\nonumber\\
=&16\pi\hbar^2t\int_0^{\infty}d\om\om^4|\alpha(\om)|^2[n^2(\om)+n(\om)].
\label{mom2}
}
\end{widetext}
In the penultimate step we have used the long-time approximation  $\int_{-\infty}^\infty \dd K \frac{\sin^2\bkt{(\om_K - \om_k)t/2}}{\sbkt{(\om_K - \om_k)/2}^2} \cong \int_{-\infty}^\infty \dd K \,2\pi t \delta (\omega_K - \omega_k)$.

In thermal equilibrium the average rate of increase of the kinetic energy must balance the average rate $\la \mc{F}v\ra$ at which the particle loses energy because of the frictional force:
\be
\Big\la\frac{d}{dt}\frac{P_y^2}{2m}+\mc{F}v\Big\ra=0,
\label{nov6}
\ee
or, from (\ref{mom6}), (\ref{mom2}), and the energy equipartition assumption $\la\frac{1}{2}mv^2\ra=\frac{1}{2}k_BT$ for thermal equilibrium at temperature $T$,
\eqn{
\frac{8\pi\hbar^2}{m}\int_0^{\infty}d\om\om^4|\alpha(\om)|^2[n^2(\om)+n(\om)]\non\\
+\frac{4\hbar k_BT}{m}\int_0^{\infty}d\om\om^3\alpha_I(\om)\frac{\pa n}{\pa\om}=0.
}
Using the relation $|\alpha(\om)|^2=\alpha_I(\om)/2\pi\om$ (Eq. (\ref{pol})), and the fact that $\alpha_I(\om)>0$ for all $\om>0$, we obtain
\be
\frac{\pa n}{\pa\om}=-\frac{\hbar}{k_BT}(n^2+n),
\label{nov67}
\ee
which is satisfied as expected by 
\be
n(\om)=\frac{1}{e^{\hbar\om/k_BT}-1}.
\ee

\section{Hyperbolic Motion in Vacuum}\label{hyperbolic}
We expect the results above to apply to the case of hyperbolic motion in the zero-temperature vacuum if we simply replace $k_BT$ by $\hbar a/2\pi$. In particular, we expect the coupling of the particle to the quantum vacuum field to cause quantum fluctuations about its classical trajectory. A calculation showing this seems nevertheless worthwhile. We now present such a calculation for the momentum fluctuations experienced by a polarizable particle in hyperbolic motion in vacuum.  

We start with Eqs. (\ref{eq2}). (\ref{eq3}), and (\ref{momm7}), but now with
\be
t(\tau)=\frac{1}{a}\sinh{a\tau} \ \ \ \ {\rm and} \ \ \ \ y(\tau)=\frac{1}{a}\cosh{a\tau}
\ee
for hyperbolic motion, and therefore
\be
\om t(\tau)+ky(\tau)=\om(t(\tau)+y(\tau))=
\frac{\om}{a}e^{a\tau}
\ee
for a particle accelerated in the $y$ direction and for modes $k=\om>0$, which we assume for now. We will carry out the calculation with (proper) time durations $\tau$ registered by a clock moving with the particle characterized by a polarizability $\alpha(\om)$. Since $\phi_0(y,t)$ may be expressed as a function of $\tau$, we write it in terms of operators $g(\Om)$ and $\gd(\Om)$ as follows:
\be
\phi_0(\tau)=\int_0^{\infty}d\Om\big[g(\Om)e^{-i\Om\tau}+\gd(\Om) e^{i\Om\tau}\big].
\label{momm12}
\ee
Then
\be
e\dot{x}=\int_0^{\infty}d\Om\Om^2\big[\alpha(\Om)g(\Om)e^{-i\Om\tau}+\alpha^*(\Om)\gd(\Om)e^{i\Om\tau}\big],
\ee
\be
\frac{\pa\phi_0}{\pa y}=i\int_0^{\infty}d\Om\Om\big[g(\Om)e^{-i\Om\tau}-\gd(\Om)e^{i\Om\tau}\big],
\ee
and
\begin{widetext}
\eqn{
\label{eq:pyt}
P_y(\tau)=\frac{i}{2}\int_0^{\infty}d\Om_1\int_0^{\infty}d\Om_2\Om_1^2\Om_2&\sbkt{\alpha(\Om_1)g(\Om_1)g(\Om_2)G_{12}(\tau)-\alpha(\Om_1)g(\Om_1)\gd (\Om_2)F_{12}(\tau)
+\alpha^*(\Om_1)\gd (\Om_1)g(\Om_2)F^*_{12}(\tau)\right.\nonumber\\
&\quad\quad\quad\quad\quad\quad\quad\quad\quad\quad\quad\quad\quad\quad\quad\quad\quad\left.\mbox{}-\alpha^*(\Om_1)\gd (\Om_1)\gd (\Om_2)G^*_{12}(\tau)}+{\rm hc},
}
\end{widetext}

where we have defined
\bea
F_{12}(\tau)&=&2e^{-i(\Om_1-\Om_2)\tau/2}\frac{\sin{\frac{1}{2}(\Om_1-\Om_2)\tau}}{\Om_1-\Om_2},\nonumber\\
G_{12}(\tau)&=&2e^{-i(\Om_1+\Om_2)\tau/2}\frac{\sin{\frac{1}{2}(\Om_1+\Om_2)\tau}}{\Om_1+\Om_2}.
\eea
Now we proceed as in the calculation for thermal equilibrium, regarding $\phi_0$ and its derivatives as Gaussian random processes. Then in taking expectation values we can treat the operators depending on $\Om_1$ and $\Om_2$ as uncorrelated.

\subsection{Momentum fluctuations}
The momentum fluctuations $\la\Delta P_y^2(\tau)\ra=\la P_y^2(\tau)\ra$ of the particle can be obtained as:
\begin{widetext}
    \eqn{
\la P_y^2(\tau)\ra=4{\rm Re}\int_0^{\infty}d\Om_1\int_0^{\infty}d\Om_2\int_0^{\infty}d\Om_3\int_0^{\infty}d\Om_4\Om_1^2\Om_3^2\Om_2\Om_4 \alpha(\Om_1)\alpha^*(\Om_3)\la g(\Om_1)\gd(\Om_3)\ra\la \gd(\Om_2)g(\Om_4)\ra F_{12}(\tau)F^*_{34}(\tau).
\label{momm20}
}
\end{widetext}

We have dropped terms involving $G_{ij}$ and $G^*_{ij}$. Because of their nonresonant character they make no contribution to $\la P_y^2\ra$. 

From (\ref{momm12}), again for a particle accelerated along the $y$ direction,
\eqn{
g(\Om)=&\frac{1}{2\pi}\int_{-\infty}^{\infty}d\tau e^{i\Om\tau}\phi_0(\tau)\\
\equiv&\frac{1}{2\pi}\sum_kC_k\sbkt{a_k\xi (\om_k,\Om)+\ad_k\eta(\om_k,\Om)},
}
where we have defined the functions $ \xi (\omega_k, \Omega)$ and $ \eta (\omega_k, \Omega)$,  corresponding to the time-dependent Doppler shift observed by the accelerated observer~\cite{AlsingMilonni2004}, as

\eqn{
\label{eq:xi}
\xi (\om,\Om)=& \int_{-\infty}^{\infty}d\tau e^{i\Om\tau}e^{i\frac{\om}{a}e^{a\tau}}\non\\
=& \frac{1}{a}\Gamma\bkt{\frac{i\Om}{a}}\bkt{\frac{\om}{a}}^{-i\Om/a}e^{-\pi\Om/2a},\\
\label{eq:eta}
\eta(\om,\Om)=&\int_{-\infty}^{\infty}d\tau e^{i\Om\tau}e^{-i\frac{\om}{a}e^{a\tau}}\non\\
=& \frac{1}{a}\Gamma\bkt{\frac{i\Om}{a}}\bkt{\frac{\om}{a}}^{-i\Om/a}e^{\pi\Om/2a},
}
where $\Gamma(x)$ is the Gamma function (see Appendix B for details). 

We must now evaluate the correlation functions appearing in (\ref{momm20}) that relate to the power spectrum of the vacuum fluctuations as seen by the accelerated particle. Consider first
\be
\la g(\Om_1)\gd(\Om_3)\ra=\bkt{\frac{1}{2\pi}}^2\sum_kC_k^2\xi(\om_k,\Om_1)\xi^*(\om_k,\Om_3).
\ee
We have used the vacuum expectation values $\la a_ka_{k\pr}\ra=\la \ad_{k}\ad_{k\pr}\ra=\la \ad_ka_{k\pr}\ra=0$ and $\la a_k\ad_{k\pr}\ra=\delta_{kk\pr}$. Now, in the usual mode continuum limit,
\eqn{
\sum_kC_k^2\xi(\om_k,\Om_1)\xi^*(\om_k,\Om_3)
=\frac{\hbar}{a^2}\Gamma\bkt{\frac{i\Om_1}{a}}\Gamma^*\bkt{\frac{i\Om_3}{a}}&\non\\
\times e^{-\pi(\Om_1+\Om_3)/2a}e^{i(\Om_1-\Om_3)/a}\int_0^{\infty}d\om\frac{1}{\om}\bkt{\frac{\om}{a}}^{i(\Om_1-\Om_3)/a}.&
}
The integral may be evaluated with the change of variable $x=\ln(\om/a)$ to
obtain $2\pi a\delta(\Om_1-\Om_3)$, and so
\be \label{eq:ggdag}
\la g(\Om_1)\gd(\Om_3)\ra=\frac{\hbar}{2\pi a}\abs{\Gamma\bkt{\frac{i\Om_1}{a}}}^2e^{-\pi\Om_1/a}\delta(\Om_1-\Om_3).
\ee
We find similarly that
\be\label{eq:gdagg}
\la \gd(\Om_2)g(\Om_4)\ra=\frac{\hbar}{2\pi a}\abs{\Gamma\bkt{\frac{i\Om_2}{a}}}^2e^{\pi\Om_2/a}\delta(\Om_2-\Om_4)
\ee
and $\la g(\Om_1)g(\Om_3)\ra=\la\gd(\Om_1)\gd(\Om_3)\ra=0$. 
After some algebra we obtain
\begin{widetext}
    \eqn{
\la P_y^2(\tau)\ra=&4\int_0^{\infty}d\Om_1\int_0^{\infty}d\Om_2\int_0^{\infty}d\Om_3\int_0^{\infty}d\Om_4\Om_1^2\Om_3^2\Om_2\Om_4\abs{\alpha(\Om_1)}^2\bkt{\frac{\hbar}{2\pi a}}^2\abs{\Gamma\bkt{\frac{i\Om_1}{a}}}^2\abs{\Gamma\bkt{\frac{i\Om_3}{a}}}^2\nonumber\\
&\times \delta(\Om_1-\Om_3)\delta(\Om_2-\Om_4)e^{-\pi(\Om_1-\Om_2)/a}\frac{\sin^2\frac{1}{2}(\Om_1-\Om_2)\tau}{\frac{1}{2}(\Om_1-\Om_2)]^2}\nonumber\\
\rightarrow&  4\bkt{\frac{\hbar}{2\pi a}}^2(2\pi\tau)\int_0^{\infty}d\Om_1\Om_1^6|\alpha(\Om_1)|^2\abs{\Gamma\bkt{\frac{i\Om_1}{a}}}^4.
}
\end{widetext}

Finally we use the identity
\be
\abs{\Gamma\bkt{\frac{i\Om}{a}}}^4=\frac{\pi^2a^2/\Om^2}{\sinh^2(\pi\Om/a)}=\frac{\pi^2a^2}{\Om^2}\frac{4e^{2\pi\Om/a}}
{(e^{2\pi\Om/a}-1)^2}
\ee
to obtain
\be\la P_y^2(\tau)\ra=16\pi\hbar^2\tau\int_0^{\infty}d\om\om^4|\alpha(\om)|^2[n^2(\om)+n(\om)],
\label{momm21}
\ee
where we have included a factor 2 to account for modes with $k<0$ as well as $k>0$, and have defined
\be 
\label{Eq:Tdu}
n(\om)=(e^{\hbar\om/k_BT}-1)^{-1}, \ \ \ \ T\equiv \hbar a/2\pi k_B.
\ee
The result (\ref{momm21}) has the same form as (\ref{mom2}), as expected. Interestingly, the Bose--Einstein factor $n^2+n$ in (\ref{mom2}) was obtained from the algebra of annihilation and creation operators, whereas the derivation of (\ref{momm21}) depended mainly on the analytical properties of $\xi (\om,\Om)$ and $\eta(\om,\Om)$, as well as the vacuum expectation values $\la a_k\ad_{k\pr}\ra=\delta_{kk\pr}$.

\subsection{Frictional force}

Finally we consider briefly the frictional force on the particle's center-of-mass, starting with

\eqn{ \mc{F} = \avg{\frac{d P _y (\tau )}{d \tau }} .
}
Using Eq.~\eqref{eq:pyt} and ignoring the non-resonant terms, this simplifies to
\eqn{
&\mc{F} = -\frac{i}{2} \int_0 ^\infty \dd\Omega_1 \int_0 ^\infty \dd\Omega_2 \Omega_1 ^2 \Omega_2\non\\
&\sbkt{ - \alpha\bkt{\Omega_1 }\avg{ g\bkt{\Omega_1 }g^\dagger\bkt{\Omega_2 } }\der{F_{12}(\tau )}{\tau }\right.\non\\
&\left.+ \alpha^\ast\bkt{\Omega_1 }\avg{ g^\dagger\bkt{\Omega_1 }g\bkt{\Omega_2 } }\der{F_{12}^\ast(\tau )}{\tau } } + hc .
}
Noting that $ \der{F_{12}(\tau )}{\tau } = e^{- i (\Omega _1 - \Omega_2 )\tau}$, and using the expectation values of the correlation functions from Eq.~\eqref{eq:ggdag} and \eqref{eq:gdagg}, yields

    \eqn{
\mc{F}  =& { \hbar }\int_0 ^\infty\dd\Omega \Omega^2 \alpha_I\bkt{\Omega } \coth \bkt{\pi \Omega/a}\\
=&2\hbar\int_0^{\infty}d\om\om^2\alpha_I(\om)\sbkt{n(\om)+\frac{1}{2}},
 }
 
where in the last step we have again defined ${n}(\om)=(e^{2\pi\om/a}-1)^{-1}$ and have expressed $ |\Gamma \bkt{\frac{i \Omega}{a}}|^2$ as $ \frac{\pi a/\Om }{\sinh (\pi \Omega /a)}$.
This expression has the form $\mc{F}=\frac{1}{2}\int_0^{\infty}R(\om)\rho(\om)$, as in Eq. 
(\ref{nov5}) but with $\rho(\om)$ appearing instead of the Lorentz-transformed $\rho\pr(\om)$.  In our heuristic approach we now argue, as in Section \ref{sec:friction}, that if the particle has a velocity $v$ the spectral density $\rho(\om)$ experienced by it should be $\rho\pr(\om)$, just as in Eq. (\ref{novv5}), such that
\be
\mc{F}=2\hbar\int_0^{\infty}d\om\om^2\alpha_I(\om)\Big[n(\gamma\om(1+v))+\frac{1}{2}\Big].
\ee
The net frictional force is then, exactly as in Section \ref{sec:friction},
\be
\mc{F}=4\hbar v\int_0^{\infty}d\om\om^3\alpha_I(\om)\frac{\pa n}{\pa\om}.
\label{nov66}
\ee
Assuming an equilibrium power balance according to Eq. (\ref{nov6}) we obtain, from (\ref{nov66}) and (\ref{momm21}), Eq. (\ref{nov67}), but now with $k_BT=\hbar a/2\pi$ if we assume the energy  equipartition relation $\la\frac{1}{2}mv^2\ra=\hbar a/2\pi$ for thermal equilibrium at the Davies-Unruh temperature.

\section{Summary}\label{summary}
The 2D scalar QED model was used previously \cite{raine,rfo1} to address the question of whether there is radiation from a uniformly accelerated oscillator in  vacuum. We have focused here not on the radiation, or lack thereof, but on the force on a polarizable particle in thermal equilibrium (at some temperature $T>0$) with the field or in uniform acceleration in a vacuum. In both cases the momentum fluctuations of the particle depend on the photon number variance having the Bose--Einstein form $n^2(\om)+n(\om)$, where $n(\om)$ is the mean photon number. In the former case this variance reflects the blackbody photon statistics of real photons, whereas in the latter case it emerges, along with the Davies-Unruh temperature, as a consequence of the different vacuum field experienced by the particle in hyperbolic motion. The  quantum-mechanical momentum fluctuations of the particle imply that only the {\it average} acceleration is constant when a constant external force acts on the particle. We also demonstrate  that the frictional force  and momentum fluctuations obtained for hyperbolic motion in vacuum are related to each other via the power balance (fluctuation-dissipation) relation (\ref{nov6}).
 
\section{Acknowledgements}Peter Milonni has fond memories of discussions with Franco Persico many years ago in Trieste and Erice, and thanks the editors for the opportunity to contribute to this special issue in his memory. Kanu Sinha's research was supported in part by the National Science Foundation under Grant No. NSF PHY-2309341 and by the John Templeton Foundation under Award No. 62422.


\appendix
\section*{Appendix A. Equations of motion}
\renewcommand{\theequation}{A.\arabic{equation}}
\setcounter{equation}{0}

The equations of motion for the charge at $Y=0$ are given by:

\eqn{\label{eq:dxdt}
\der{x}{t} =& \frac{1}{2m }\bkt{p_x - e\phi (0,t)},\\
\der{p_x}{t} = &- m \omega_0 ^2 x,
}
such that 
\eqn{
\der{^2x}{t^2} + \omega_0 ^2 x  = - \frac{e}{m} \pard{\phi (0,t) }{t}.
}
Dividing the field into source-free $(\phi_0)$ and radiation reaction parts $(\phi_{RR})$, we get:
\eqn{\label{eq:d2xdt2}
\der{^2x}{t^2} + \omega_0 ^2 x  +  \frac{e}{m} \pard{\phi_{RR}(0,t)}{t}= - \frac{e}{m} \pard{\phi_0(0,t)}{t},
}
where a friction term arises explicitly from the radiation reaction part. 

The Heisenberg equation of motion for the field operator $ a_k$ is 

\eqn{
&i \hbar \der{a_k}{t} = \sbkt{a_k, \sum _{k'} \hbar\omega_k a_{k'}^\dagger a_{k'} - \frac{e}{m} \phi(0,t) p_x  +\frac{e^2}{m} \phi^2(0,t)} \\
&\implies \der{a_k}{t} = - i \omega_k a_k + i \frac{e}{\hbar} C_k \bkt{\frac{p_x - e\phi (0,t)}{m}}
}
Using Eq.~\eqref{eq:dxdt} in the above and integrating over time, we obtain:

\eqn{a_k (t)  = a_k (0) e^{-i \omega_k t } + \frac{i e }{\hbar } C_k  \int_0 ^t \dd\tau \dot{x}(\tau ) e^{i \omega_k \bkt{t - \tau}}
}
Substituting the above in Eq.~\eqref{eq:phiyt}, one gets the free and source parts of the field as follows
\eqn{
\phi(y,t) =&\underbrace{\sum_k C_k \bkt{a_k (0) e^{- i \omega_k t } e^{i k y} + a_k ^\dagger (0) e^{i \omega_k t} e^{-iky} }}_{\phi_0 (y,t) \ \text{(Free field)}} \non\\
& + \underbrace{\sum_k \frac{2 C_k^2  e}{\hbar } \int _0 ^t \dd\tau\dot{x}(\tau ) \sin \bkt{t - \tau }}_{\text{Radiation reaction}}  
}

The radiation reaction part of the field as it appears in the equation of motion for the charge is

\eqn{
\pard{\phi_{RR}(0,t)}{t} = &\frac{2e}{\hbar} \sum_k C_k^2 \omega_k \int_0 ^t \dd\tau \dot{x}\bkt{\tau } \cos \bkt{\omega_k (t - \tau)}\non\\
& \rightarrow2 \pi e \dot{x}\bkt{t },
}
where in the last step we have taken the continuum limit of the summation ($ \sum_k\rightarrow\frac{L}{2\pi }\int\dd k) $. Substitution of the above in Eq.~\eqref{eq:d2xdt2} yields Eq.~\eqref{eq1}.

\section*{Appendix B. Derivation of $\xi (\omega_k,\Omega)$ and  $ \eta (\omega_k,\Omega)$}
\renewcommand{\theequation}{B.\arabic{equation}}
\setcounter{equation}{0}
In this appendix we present a derivation of the functions $ \xi (\omega, \Omega)$ and $ \eta (\omega , \Omega )$ that are related to the frequency spectrum seen by the accelerated observer~\cite{AlsingMilonni2004}. We define the variable $ z = e^{a\tau }$, such that, $ \dd z = a e^{a \tau} \dd\tau $, yielding:
\eqn{ 
\xi (\om,\Om)=&\int_{-\infty}^{\infty}d\tau e^{i\Om\tau}e^{i\frac{\om}{a}e^{a\tau}}\non\\
=& \frac{1}{a}\int_0 ^\infty \dd z\,z^{i \Omega/a - 1} e^{- z/(i a /\omega)} 
}
Substituting  $ \tilde z = z/(i a /\omega)$, we obtain
\eqn{ 
\xi (\om,\Om)=& \frac{1}{a}\bkt{\frac{i a}{\omega}}^{i \Omega/a}\int_0 ^\infty \dd \tilde z\,\tilde z^{i \Omega/a - 1} e^{- \tilde z} \non\\
=& \frac{1}{a} \Gamma\bkt{\frac{i \Omega}{a}}\bkt{\frac{\omega}{a}}^{-i \Omega/a} \bkt{e^{i  \pi/2} } ^{i \Omega /a},
}
which readily gives Eq.~\eqref{eq:xi}. Similarly,
\eqn{\eta(\omega, \Omega) =& \int_{-\infty}^{\infty}d\tau e^{i\Om\tau}e^{- i\frac{\om}{a}e^{a\tau}} \non\\
= & \frac{1}{a}\Gamma\bkt{\frac{i\Om}{a}}\bkt{-\frac{\om}{a}}^{-i\Om/a}e^{-\pi\Om/2a} \non\\
=& \frac{1}{a}\Gamma\bkt{\frac{i\Om}{a}}\bkt{e^{i\pi}\frac{\om}{a}}^{-i\Om/a}e^{-\pi\Om/2a},
}
which yields Eq.~\eqref{eq:eta}.

\end{document}